# Uncovering the properties of homo-epitaxial GaN devices through cross-sectional infrared nanoscopy

Hossein Zandipour[1], Felix Kaps[2], Robin Buschbeck[2], Maximilian Obst[3], Aditha Senarath[3], Richarda Niemann[3], Niclas S. Mueller[4,5], Gonzalo Alvarez-Perez[6], Katja Diaz-Granados[3], Ryan A Kowalski[3], Jakob Wetzel[2], Raghunandan Balasubramanyam Iyer[1], Matthew Wortel[1], J. Michael Klopf[7], Travis Anderson[8], Alan Jacobs[9], Mona Ebrish[3], Lukas M. Eng[2,10], Alexander Paarman[4], Susanne C. Kehr[2], Joshua D. Caldwell[3+], Thomas G. Folland [1]*

[1] Department of Physics and Astronomy, University of Iowa, Iowa City, IA 52242, USA.

[2] Institute of Applied Physics, Technische Universität Dresden, 01062 Dresden, Germany.

[3] Vanderbilt University, Nashville, TN 37235, USA.

[4] Department of Physical Chemistry, Fritz Haber Institute of the Max Planck Society, Berlin, Germany.

[5] Department of Physics, Freie Universität Berlin, Berlin, Germany.

[6] Center for Biomolecular Nanotechnologies, Italian Institute of Technology, Lecce, Italy.

[7] Helmholtz-Zentrum Dresden-Rossendorf, 01328 Dresden, Germany.

[8] University of Florida, Gainesville, FL 32611, USA.

[9] U.S. Naval Research Laboratory, Washington, DC 20375, USA.

[10] ctd.qmat - Excellence Cluster TU Dresden-Würzburg, Dresden, Germany.

*thomas-folland@uiowa.edu

+josh.caldwell@vanderbilt.edu

## Abstract:

Validating material performance in electrical devices is crucial to product development. For Gallium Nitride (GaN) devices, evaluating material factors such as defects, dopant concentration, and overall production quality is essential to ensure their performance in advanced electronic and optoelectronic applications. This work demonstrates that scattering-type scanning near-field optical microscopy (s-SNOM) can meet the demanding performance requirements for characterizing homoepitaxial GaN devices. Specifically, we show that combining s-SNOM results in the mid-IR and terahertz (THz) spectral ranges can disentangle carrier and lattice changes in a GaN p-i-n diode, which is not possible using one spectral range alone. We observe strong, resonant near-field signals near the LO phonon mode of GaN that correlate well with point-dipole models. This data shows great sensitivity to the local carrier density, with changes on the order of $10^{18}$ cm$^{-3}$ easily resolved experimentally. Further, we demonstrate high sensitivity to sub-surface defects, which remain a significant challenge for other non-destructive techniques. To validate the power of s-SNOM imaging, our results are compared to traditional metrologies, including micro-Raman mapping and Kelvin Probe Force Microscopy (KPFM). Our results show that s-SNOM shows superior resolution and sensitivity to perturbations, highlighting the power of this technique in semiconductor device characterization.

## 1. Introduction

As the demand for high-performance electronic and optoelectronic devices has risen, developing mature wide band gap (WBG) semiconductors such as gallium nitride (GaN) has become increasingly important. GaN, with its direct bandgap of ~3.4 eV, high breakdown voltage, good thermal conductivity, and high electron mobility, has been demonstrated to enable the fabrication of efficient power devices[1,2]. This has included high electron mobility transistors (HEMT), and high-frequency RF components that outperform traditional silicon-based technologies [3,4]. Modern GaN devices offer substantial advantages in efficiency and performance compared to traditional semiconductor technologies[5–9]. To push GaN to its physical limit, homoepitaxial growth of GaN devices structures on native substrates has been proposed as a route to reduce dislocation densities, leading to improved reliability and efficiency in next-generation devices[10–13]. However, the growth of GaN substrates and associated epilayers is significantly less developed than more mature platforms such as Silicon Carbide, resulting in defects and variations that can render device performance inconsistent or unreliable[14]. As such, a greater understanding of material properties in WBG semiconductors is critical[15–20].

Techniques such as X-ray topography, electron microscopy, atomic force microscopy (AFM), and Raman spectroscopy are often employed to analyze the physical structure, crystal quality, carrier density, and surface morphology of the device[21–24]. X-ray topography, while capable of measuring precise changes in the lattice, lacks sufficient spatial resolution to reliably detect individual point and line defects like dislocations in GaN, and cannot quantify free carrier information. Electron imaging techniques such as electron channeling contrast imaging and transmission electron microscopy can visualize point defects at the nanoscale. However, they provide poor contrast for doping-related variations and often require specialized sample preparation techniques for high-quality images[25–30]. Atomic force microscopy provides surface level topographic and some doping information but does not offer insights into crystal structure[31]. Alternatively, Kelvin Probe Force Microscopy (KPFM) provides high-resolution maps of the local contact potential difference, revealing band-bending and work-function contrasts associated with doping[32–35]. However, it lacks information on local strain distributions. Finally, optical methods such as Raman spectroscopy can in principle image both carrier density and lattice properties. Thus, Raman is widely used to probe doping and strain through the longitudinal optical phonon plasmon coupling (LOPC) effect, enabling the extraction of the carrier concentration via dielectric modeling, but this method is inherently diffraction-limited[36–41]. Conventional Raman spectroscopy faces additional challenges including depth sensitivity, which reduces the accuracy of analysis[42]. While confocal Raman microscopy can improve vertical resolution, it is still typically limited to a few microns in GaN, *i.e.*, thicker than many device layers. This lack of surface sensitivity can be addressed by directly measuring the properties of GaN materials at infrared and THz wavelengths, where it is possible to track the energy and damping of lattice vibrations and the free electron plasma[43]. However, the diffraction-limited nature of conventional infrared spectroscopy limits its ability to resolve local changes in properties.

An emerging technique for measuring the performance of electronic devices is scattering-type scanning near-field optical microscopy (s-SNOM)[44–46]. s-SNOM combines an AFM with optical illumination to produce high spatial resolution images of the optical properties of a surface, and hence the properties of the associated materials[47–51]. This is achieved by exploiting the evanescent waves generated when light interacts with the sharp AFM tip, resulting in imaging at a spatial resolution dictated by the radius of curvature of the tip (~20 nm), which is orders of magnitude smaller than the far-field wavelength[52]. Given the long wavelength of infrared and THz light, this offers many orders of magnitude improvement in spatial resolution[53]. s-SNOM has already enabled studies of a diverse range of materials by leveraging broadly tunable laser sources[54–58]. This has included analysis of semiconductor devices and junctions, where monitoring doping and strain is required for functionality[59–62]. One challenge in performing such studies is that often imaging over a narrow spectral range presents challenges for gaining device-relevant insights.

Choosing frequencies between THz and mid-infrared (MIR) imaging is particularly critical[63]. This becomes even more complex for doped polar semiconductors, where the free electron plasma, transverse optical (TO) phonons and longitudinal optical (LO) phonons all contribute to the optical response and provide complementary information. Unpicking changes in the lattice (which influence phonon properties) and carrier density therefore becomes a key challenge in materials such as SiC and GaN. Furthermore, s-SNOM is typically only sensitive to the top layers of a surface, making analysis of buried heterostructures almost impossible.

In this study, we demonstrate that THz through MIR s-SNOM can provide the contrast and high spatial resolution to characterize GaN device power electronic devices. In particular, we show that we can resolve such information in cross-section, extracting changes in the carrier density as a function of depth into the substrate. This is critical for characterizing vertical power devices, where device layers are buried in the substrate. By performing single-frequency imaging across a broad spectral range, we can then measure local variations in carrier density and the lattice. In such experiments, changes in scattering in the THz regime relate purely to carrier concentration. In contrast, in the MIR they relate to changes in both the lattice properties and carrier concentration, through the LOPC effect. The observed changes in all images correlate well with what is expected from a simple point-dipole model (PDM) for the MIR and THz near-field response. Variations in carrier density are seen using both THz and MIR excitations, with the LOPC driven interaction in the mid-infrared offering higher contrast imaging and high signal-to-noise levels. However, comparing THz and MIR images, additional changes in scattering are observed that cannot be explained through carrier-density variations alone. Through this comparison, we can attribute these changes to local lattice variations, between epitaxially grown layers and the substrate. This demonstrates that the combination of MIR and THz imaging can uncover material behaviors which cannot be seen by one of these approaches alone. To validate our results, we further employ Micro-Raman mapping and KPFM on the same sample. We find broad agreement between these experiments. However, both Raman and KPFM measurements are unable to resolve localized defects, demonstrating the power of s-SNOM in addressing these specific materials characterization challenges. These results and comparisons offer a generalizable framework to other polar wide gap materials, such as Aluminum Nitride and Gallium Oxide.

## 2. Experimental Design and Setup

The device was obtained from a previous study on edge-termination design for vertical GaN PIN diodes [64]. The device is grown on an N-type GaN substrate with nominal dopant concentration of $3 \times 10^{18} cm^{-3}$. The epitaxial layers grown on top of the substrate are a 0.5μm highly doped region (n$^{++}$), an 8μm drift layer, and a 0.5μm Mg compensated P-type GaN. The dopant concentrations of the drift layer and the compensated layer on top are $1.6 \times 10^{16} \ cm^{-3}$ and $3 \times 10^{17} cm^{-3}$, respectively. The chip consisted of an array of high voltage PIN devices fabricated on a single chip. To prepare the sample, the chip was first held vertically in a cast using metal clips and then entirely embedded in epoxy using Epofix resin and hardener. After curing, the epoxy was carefully cross-sectioned through the chip using a diamond saw to be able to access the layers from the side. Finally, the cross-sectioned surface underwent a multistage mechanical polishing. The process began with multistage wet sanding, starting from 400 Grit to 1200 Grit on a plate glass, and then finished up by the final polish using MasterPrep 0.05μm Alumina solution to achieve a smooth surface roughness for high-quality imaging. Figure 1a represents the final prepared sample showing the cross-sectioned chip in the epoxy, which was held via a metal clamp. The bottom section of the figure represents a zoomed-in optical image of the sample. It shows the optically smooth surface of the sample with different sections (epoxy, metal clamp, gold contact and the device itself) noted. In addition, an AFM scan of the device at the edge in Figure 1b. shows a topographic variation of less than a micron, and RMS roughness of approximately 0.36μm alongside a visualization of the expected structure on the scan. This suggests a high-quality optical polish comparable to that used on epi-ready substrates, which is appropriate for optical techniques such as s-SNOM.

To differentiate between device layers, it is crucial to choose probe wavelengths where there is measurable contrast between device layers. To identify the relevant frequencies for the s-SNOM scans, we employed a simplified PDM of tip-sample interaction to provide a semi-quantitative framework for interpreting the change in carrier density with respect to the epitaxial layer. The model uses the near-field interaction between the AFM tip and sample using the point dipole approximation, where the effective polarizability of the system is given by[53–55,65]:

$$\alpha_{effective} = \frac{\alpha_{tip}}{1 - \frac{\alpha_{tip} \cdot \beta}{16\pi(r+z)^3}} \tag{1}$$

where $\alpha_{tip}$ is the tip polarizability in vacuum and calculated as: $\alpha_{tip} = 4\pi r^3 \frac{(\varepsilon_{tip}-1)}{(\varepsilon_{tip}+2)}$, when $\varepsilon_{tip}$ is the permittivity of the AFM tip, $r$ is the tip's radius and $z$ is the tip-sample distance that is modulated at a certain frequency. $\beta$ is the complex reflection coefficient obtained from the dielectric constant and calculated as $\beta = \frac{\varepsilon-1}{\varepsilon+1}$ with $\varepsilon$ being the permittivity of the sample. During a full cycle of tapping the z is changing as:

$$z(t) = z_0 + A[1 - cos\varphi(t)] \tag{2}$$

where $z_0$ is the distance of tip from the sample at rest (with no mechanical oscillation), A is the tapping amplitude and $\varphi(t)$ is the tapping phase. Consequently, the $\alpha_{effective}$ varies as the tip oscillates that causes the near-field contribution to be modulated at the tapping frequency. To find the demodulated n$^{th}$ harmonics of the s-SNOM signal ($S_n$), one can simply take the n$^{th}$ Fourier component of the $\alpha_{effective}$ as follows:

$$S_n = \frac{1}{2\pi}\int_0^{2\pi} \alpha_{effective}(\varphi) \cdot e^{-in\varphi} d\varphi \tag{3}$$

To calculate the permittivity, we utilized a combination of TO-LO model for GaN and a Drude model to account for both lattice vibrations and carrier densities as shown (Refer to Figure S1 in Supplemental materials):

$$\varepsilon(\omega) = \varepsilon_\infty \frac{\omega^2 - \omega_{LO}^2 + i\gamma_{LO}\omega}{\omega^2 - \omega_{TO}^2 + i\gamma_{TO}\omega} - \frac{\omega_p^2}{\omega^2 + i\gamma_p \cdot \omega} \tag{4}$$

Here $\omega$, $\omega_p$, and $\gamma_p$ are the excitation, plasma frequencies, and plasma damping, respectively, and the TO-LO values including $\varepsilon_\infty, \omega_{LO}, \gamma_{LO}, \omega_{TO}$, and $\gamma_{TO}$ were chosen based on the available literature on GaN[43]. It is worth noting that for simplicity, the in- and out-of-plane components of the dielectric function were weighted equally to obtain an effective permittivity used in the analysis. The primary driver in contrast between layers is the change in relative dopant concentration as described by the plasma frequency:

$$\omega_p = \sqrt{\frac{n_e e^2}{\varepsilon_0 m^*}} \tag{5}$$

where $e$ is the elementary charge, $\varepsilon_0$ is the permittivity of the free space, $m^*$ is the effective mass of the carrier, and $n_e$ is the carrier concentration, which is directly proportional to the dopant concentration. It is also possible that lattice deformations could cause small variations in the phonon energies, which will shift resonances associated with the LO phonon, and need to be deconvolved from carrier density variations. The results of the PDM calculations (simulation parameters are noted in Supplementary Materials) for different dopant densities in GaN are shown in Figure 1c, revealing two main ranges where contrast can be detected (refer to Figure 1d., and 1e.). In the MIR, there is a strong resonant peak associated with the zero crossing of the dielectric function near the LO phonon energy, where scattering will be extreme. This peak is highly sensitive to relatively small changes in carrier density thanks to longitudinal optical phonon-plasmon coupling, with changes on the order of $10^{17}$ cm$^{-3}$ carriers leading to an observable change. This peak blueshifts with doping, and for the carrier densities present in the device (on the order of $10^{18}$ cm$^{-3}$) the optimal scattering occurs at 710 cm$^{-1}$. Conversely, no change in contrast is seen close to the TO phonon, which is due to the lack of near-field resonances at this frequency range. In the THz on the other hand, the change in contrast is much weaker in the s-SNOM imaging, which can be attributed to the higher losses associated with the free electron plasma. The model tells us that measurements between 100 and 350 cm$^{-1}$ are optimal to identify the changes in carrier density for this device (see Figure 1d.). Whilst the THz range offers relatively weaker contrast, it is worth highlighting that any changes in the phonon energies will also change scattering, which can cause significant contrast in the mid-infrared as has been observed previously [30]. As such, by performing multi-spectral imaging, we can disentangle these two effects.

To perform THz through MIR s-SNOM measurements on the sample, we utilize a Neaspec s-SNOM setup integrated with the free-electron laser FELBE at the Helmholtz-Zentrum Dresden-Rossendorf, Germany[42,66]. This laser source has the broad tunability capable of addressing both THz and MIR imaging using the same laser system. An AFM tip oscillates at its resonant frequency (approximately at 256 kHz) near the sample surface while being excited by a free-electron laser (FEL) tunable from 40 cm$^{−1}$ to 2000 cm$^{-1}$. The AFM tip concentrates the electric fields at its sharp tapping point interacting with the sample and alters the total scattered near-field signal. Both the far-field and near-field signals are captured using a liquid-nitrogen-cooled mercury-cadmium-telluride photoconductive detector (Judson Technologies LLC, Model J15D16 with a thallium bromo-iodide window) or a liquid helium-cooled gallium-doped germanium photoconductive detector (QMC Instruments Ltd). To determine the information embedded in the near-field scattered signal, the system demodulates the captured signal at different harmonics (usually higher harmonics) of the AFM lever's fundamental oscillation frequency. We use the third harmonic throughout this paper, which offers excellent surface sensitivity but without compromising signal-to-noise, as is seen at higher harmonics (refer to Figure S3, S4, and S5 in the supplementary material for the second harmonic results)[42,47,48]. All measurements were conducted in a self-homodyne detection scheme, where the FEL is mixed with the scattered light on the detector to increase signal-to-noise. However, this results in several drawbacks, such as mixing of amplitude and phase information, more extensively discussed in prior work[54,67].

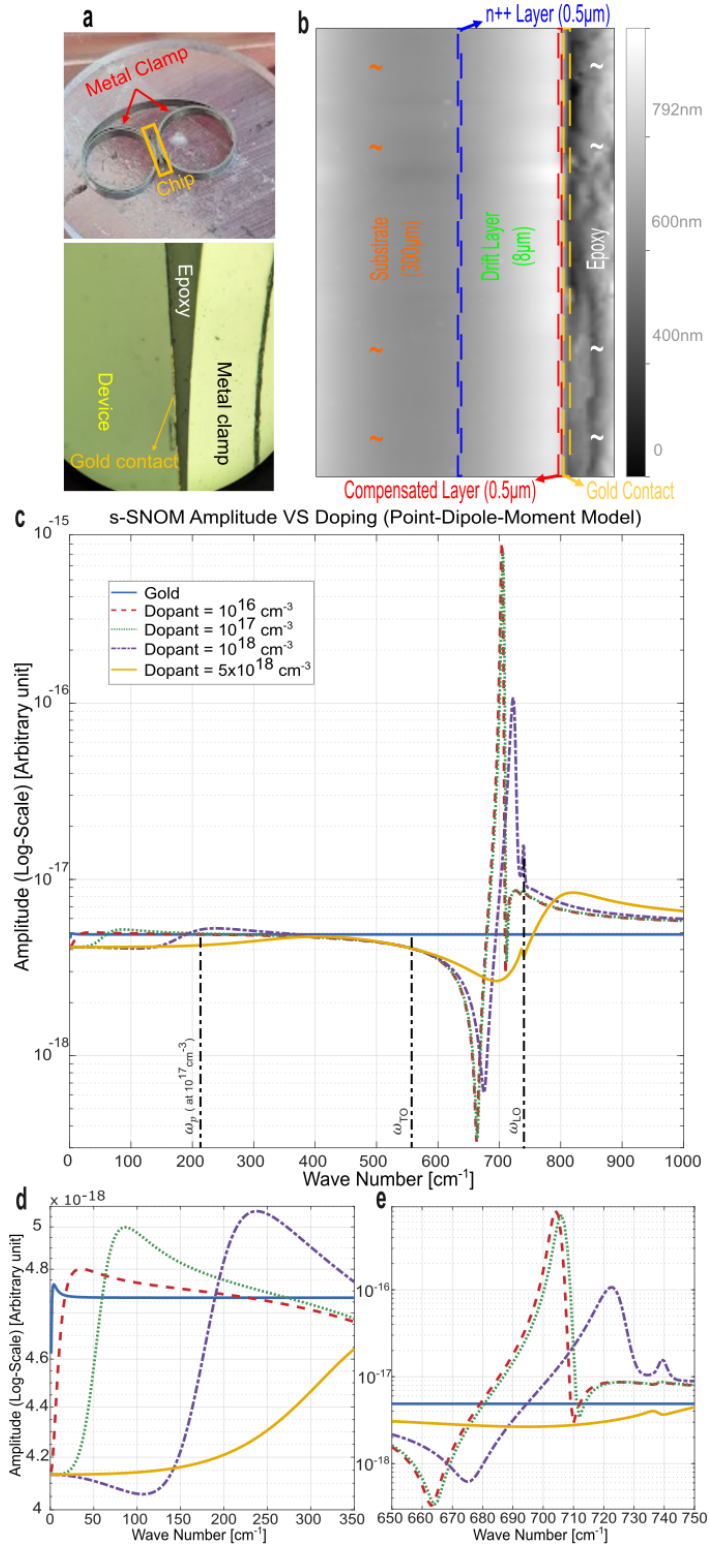

Figure 1: Overview of the prepared GaN sample, its surface morphology, and the corresponding Point Dipole Model (PDM) simulation. a) Finished sample, the chip is held by a metal clamp embedded in transparent epoxy. The bottom image shows a zoomed-in microscope image of the sample. b) AFM scan of the sample covering the area of 20μmX30μm. This shows a sub-micron step from the chip to epoxy. The approximate locations of various device layers are also labeled on this scan. c) Simulation results of the amplitude of s-SNOM signal demodulated at third harmonics using PDM (i.e. $\alpha_{effective}$ demodulated at the third harmonics over a full cycle) on GaN for different dopant concentration as well as the gold. d) The results of part c, plotted for the range of 0 cm$^{-1}$ to 350 cm$^{-1}$ highlighting the variation in the THz region. e) The results of part c, plotted for the range of 650 cm$^{-1}$ to 750 cm$^{-1}$ highlighting the variation in the MIR region.

Subsequent Raman spectroscopy was conducted on the sample in a region surrounding the area analyzed for the s-SNOM. Raman spectroscopy setup was a LabRAM HR Evolution Raman microscope (HORIBA Jobin Yvon GmbH, Oberusel, Germany). Measurements were performed in backscattering geometry using a monochromatic, linearly polarized, continuous wave, frequency-doubled Nd:YAG laser at 532 nm (100 mW, diode-pumped laser, Coherent) and a Nikon TU Plan Fluor EPI P 100× microscope objective (NA = 0.9). A motorized, rotatable half-wave plate controlled the polarization direction of the incident light, while a manually rotatable linear polarization filter controlled the polarization of the detected signal. Both elements could optionally be removed from the beam path. A 1800 l/mm grating and a Syncerity charge-coupled device (HORIBA Jobin Yvon GmbH, Oberursel, Germany) were used in the detection system. This detector setup allows for a spectral resolution of 0.013nm (0.48-0.60 cm$^{-1}$). The sample was mounted on an XY translation stage, allowing for raster scanning with a lateral resolution of approximately 0.5 μm. The lateral step size for scans can be freely chosen in 0.1 μm steps down to 0.1 μm. The one- and two-dimensional scans presented here were measured with a step size of 0.2 μm.

All KPFM measurements were performed using an attocube scanning probe microscope (*neaSCOPE*, attocube systems GmbH, Haar, Germany) system with ARROW-EFM tips optimized for surface potential measurements. The tip oscillated at a tapping amplitude of 50 nm at the first-order mechanical resonance of the cantilever Ω. Moreover, for the applied heterodyne KPFM mode, an AC voltage with an amplitude of 2.0 V and a frequency matching the second-order mechanical resonance of the cantilever is applied to the tip, while the sample is grounded at its gold contacts. Via a PI controller, the resulting electrostatic force between tip and sample is minimized by applying a DC voltage to the tip, being a direct measure of the local surface potential[35]. All shown data was obtained using an integration time of 100 ms per pixel. The KPFM measurement was performed approximately 20 μm below the previous scans, due to damage to the sample due to Raman mapping.

## 3. Experimental Results
### 3.1. s-SNOM Results

The s-SNOM imaging was performed at various frequencies and demodulated at the third harmonic over an area of 20x30 μm near the edge of one of the Au contacts. Images of the sample taken at wavelengths from THz to MIR are shown in Figure 2, showing dramatic changes in near-field signals across the spectral range, which can be interpreted in the context of the PDM. Note that this figure indicates the s-SNOM amplitude with the phase correction, and is normalized with reference to the maximum signal in the image (refer to Figure S2 for only the amplitude component of the signal). Starting with the lowest wavenumber in the Terahertz regime (≈115 cm$^{-1}$), the scans show mainly broad structural features and some contrast between the device layers, yet with modest near-field sensitivity. The substrate appears darker than the drift layer in agreement with the PDM, owing to the elevated carrier density of the substrate. At higher frequencies in the THz (≈288 cm$^{-1}$), the contrast between substrate and drift layer is reduced even further. This is due to the carrier density of both layers being lower than the n++ interface region, which leads to moderate fluctuation in scattering according to the point dipole model. However, we now see a signature of the n++ region at the interface between the substrate and the drift layer. This is consistent with a highly doped layer at the interface between the drift layer and substrate, as grown in the original sample. Close to the TO phonon of GaN (≈535 cm$^{-1}$) the device shows almost no distinguishable near-field contrast, as there is no resonant near-field signal near the TO phonon. This is in stark contrast to far-field infrared measurements, where strong reflections are seen close to the TO phonon energy[43].

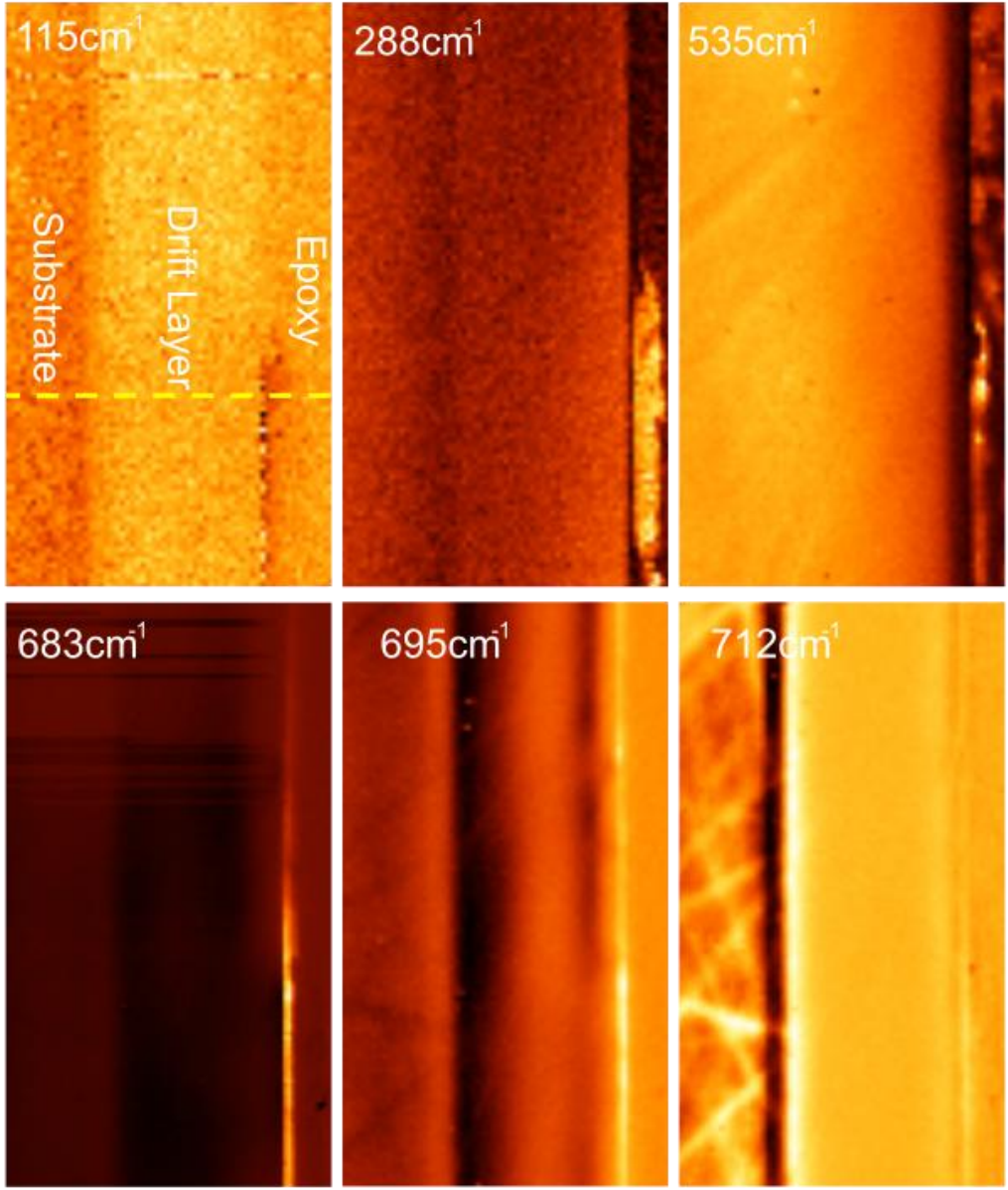

*Figure 2: The s-SNOM scans demodulated at the third harmonics at various wavenumber. This shows the amplitude of the s-SNOM with the addition of phase correction (i.e. $Amplitude \times Cos[Phase]$). All images were cropped to cover the similar scan area of approximately 17µmX30µm. The range of scans varies from 115 cm$^{-1}$ to 712 cm$^{-1}$. Note that dark and bright contrasts indicate the minimum and maximum signal strength respectively.*

In the mid-infrared, contrast reappears at 683 cm$^{-1}$ as we approach the resonant near-field signal. We see a distinction between the drift layer and substrate. The epitaxial drift layer appears uniform with the expected width of 8 µm. At 695 cm$^{-1}$ the s-SNOM signal exhibits strong, well-defined contrast between the principal device layers including substrate, drift layer, and their interface. At this frequency (see the Figure 1e.), PDM simulation points to a substantial difference between highly (n++) and lightly doped (drift) regions, and a modest difference between highly (n++) and moderately doped (substrate). Across the drift layer, we observe a change in contrast that is likely associated not with a change in optical properties, but instead with a localized surface phonon-polariton resonance across the active area [68]. Surface phonon polaritons result in bright and dark fringes on the surface due to scattering of surface waves and have been observed previously for SiC and GaN [45,68]. The narrow, bright stripe near the top of the sample (right side of the image) is likely due to scattering from the device edge.

Finally, at 712 cm$^{-1}$, the map not only fully resolves the layer structure but also reveals clear substrate non-uniformities and localized defect-like features (refer to Figure S6. in supplementary material with the features highlighted). The contrast between the highly (n++) and moderately doped region (substrate) is much larger here with a dark band present over the n++ layer, agreeing with PDM. These data sets are also broadly consistent with the second-harmonic scattering signals, which are observable in Figure S3-S5. of the supplementary materials section. In several of these, the optical contrast is flipped, in agreement with PDM calculations. However, there are several other irregularities in the 712cm$^{-1}$ image owing to the increased contrast. As highlighted in Figure S6., these manifest as 'line' like features in the substrate, as well as a bright fringe that appears just above the n++ layer. First, we rule out surface scratches as a source of the line-like features, as they are not continuous across the interfaces in the sample and are confined only to the substrate. We can gain greater insight by comparing the near-field signals between the spectral ranges studied in this image. The line features are also observable in the 695cm$^{-1}$ image, suggesting a change to the dielectric function close to the LO phonon resonance. They are also observable as very thin features at 115cm$^{-1}$ demodulated at the second harmonics in Fig S4. This suggests localized line-like defects in the substrate, which reduce the carrier density. However, the mid infrared data at 695cm$^{-1}$ exhibits lower signal in the presence of the line defects, suggesting an *increased* carrier density. This apparent contradiction can be resolved by the presence of lattice distortion, which will change the infrared response. Further, the features appear significantly wider in the mid-infrared, suggesting extended strain fields. Similar effects have been seen in silicon carbide when imaged only in the mid-infrared [69–71], but in this case it was not possible to tell if this is due to local variations in the carrier density or attributed to other changes in the lattice. The addition of a bright fringe above the n++ in the mid infrared also is suggestive of lattice deformation, given the lack of such a fringe in the THz spectrum. The combination THz and mid-infrared is crucial to make this distinction, and a key advantage of this broadband approach for semiconductor characterization.

### 3.2. Comparisons Between s-SNOM and Alternative Methods

To complement the s-SNOM analysis and validate our results, additional measurements were carried out using high resolution micro-Raman spectroscopy and KPFM. Illustrative Raman spectra from different points in the sample are shown in Figure 3a. The scans are taken at two different positions, one in the substrate and one in the drift region (marked with orange and blue in the optical image of Figure 3b.). The Raman spectra from both the substrate and the drift region are dominated by three strong features in the ~530–566 cm$^{-1}$ range, with the most intense peak near 566 cm$^{-1}$ representing the $E_2$(high) phonon, followed by peaks at approximately 556 cm$^{-1}$ and 530 cm$^{-1}$ hinting to the $E_1$(TO) and $A_1$(TO) phonons, which are close to the reported values in the literature [72]. There is minimal variation in these properties across the image, as is typical for Raman spectroscopy of GaN devices. Greater information can be gathered from the LO phonon modes around 750 cm$^{-1}$, due to the phonon plasmon coupling effect. Setting these modes aside, in the amplitude-selected Raman spectra in Figure 3a, the drift-region spectrum shows a pronounced peak at around 740 cm$^{-1}$. This high-frequency peak lies close to two LO-like modes [$A_1$(LO) and $E_1$(LO)] that have been observed in GaN[73]. As the sample geometry is in cross section, this peak likely contains contributions from both modes (located at ~730cm$^{-1}$ and ~740cm$^{-1}$), making further analysis of this peak for doping density challenging. This is because for light doping there is almost no change in the carrier density. For the substrate, this peak instead appears at around 785 cm$^{-1}$, manifesting as a bump in its spectra. This tuning of the mode can be associated with plasmon-phonon coupling, and the doping of the substrate can be found by finding where $\varepsilon(\omega) = 0$[74] in equation 4, revealing a carrier concentration estimate of 1.7 × $10^{18}$ cm$^{-3}$. The substrate carrier density is consistent with our s-SNOM results broadly, but the width of the peak and mixing of in-plane and out-of-plane phonons limits our ability to precisely resolve the carrier density[75]. By leveraging the position of these peaks to indicate doping, we also performed Raman mapping, integrating the signal over the spectral windows 720–760 cm$^{-1}$ and 760–950 cm$^{-1}$ shown in Figure 3b. Both maps show a distinct spatial contrast between the substrate (at the left side of the map) and the epitaxial drift layer, which is consistent with the layered doping structure of the sample. We also observe the presence of the heavily doped layer in this map, indicated as a line going vertically down the image. Additionally, polarized Raman line scans (refer to Figure S7 in supplementary material) were taken along the yellow dashed line (shown in Figure 3b) with parallel and perpendicular polarization with respect to the scan direction. In both cases, a clear peak appears near 740 cm$^{-1}$ in the drift layer, though it is notably suppressed when the polarization is perpendicular to the scan direction. In general, the maps broadly agree between s-SNOM and Raman, but the lower spatial resolution of Raman spectroscopy and its broad features limit our ability to resolve the fine spatial features such as the defects in the substrate.

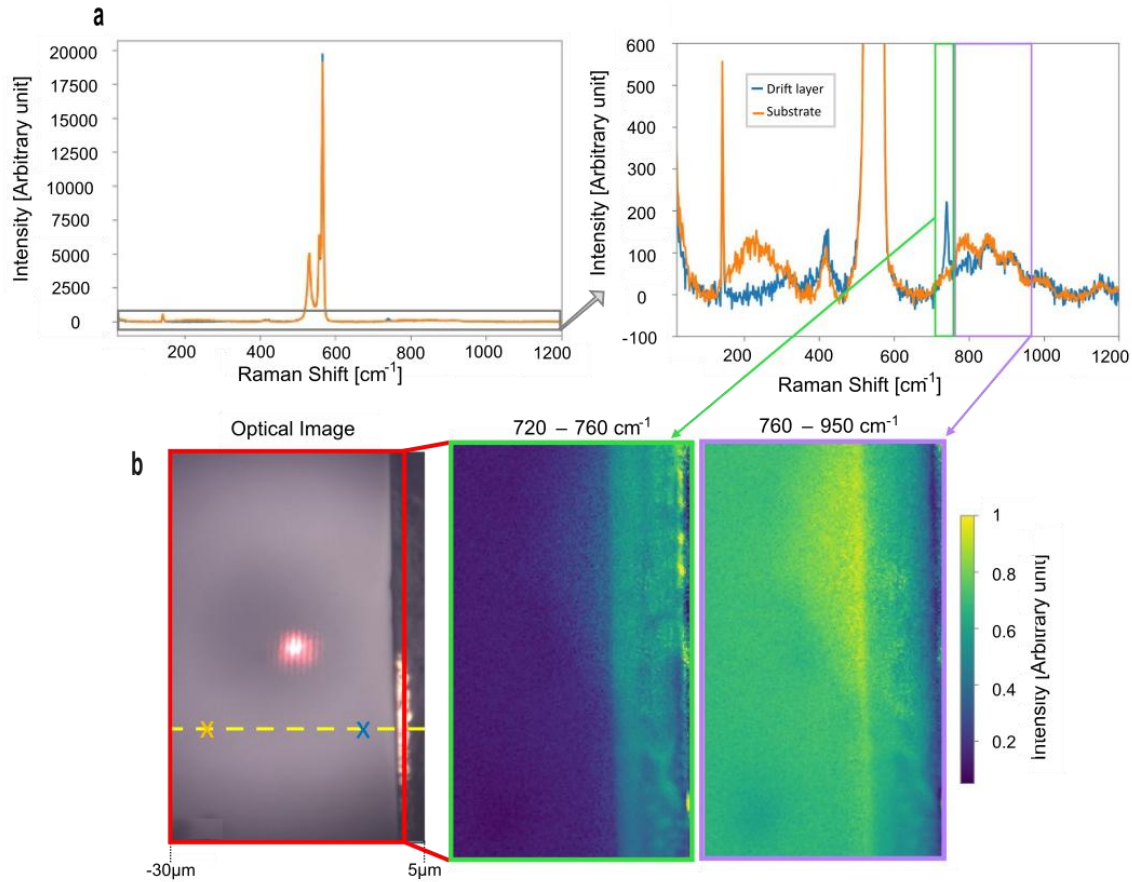

*Figure 3: Results of Raman measurement. a) The full spectrum of Raman shift at a point in substrate and the drift region (marked with orange and blue in the optical image of part b, alongside with the zoomed-in image. b) a 2D map of the scanned area (red box in the optical image) in two different spectral windows (green and purple indicated at part a) showing a difference in contrast between the drift and substrate region. The yellow dashed line in the optical image represents the approximate position of the line scan discussed in Figure 4. which follows the position indicated at the bottom*

Finally, to compare with electrical scanning probe techniques, we performed KPFM mapping as shown in Figure 4a. Note that the scan was limited to a smaller region to avoid contamination which could lead to any artificial features. As shown in Figure 4a, the KPFM map reveals contrast between the substrate, heavily doped layer and the drift layer. This is due to the change in surface potential with doping, consistent with the s-SNOM results, again validating our analysis. However, again, KPFM does not resolve any spatially distinct defects in the substrate, or dramatic changes in contrast near to the heavily doped layer. We attribute this as further confirmation that several of the bright features seen in the s-SNOM at 712 cm$^{-1}$ are due to changes in local strain. While the resolution of this imaging technique is comparable to s-SNOM, these results highlight that with optical techniques we can access additional information such as the doping distribution and irregularities about the samples such as point and line defects.

To have a more comprehensive comparison of the characterization methods discussed so far, Figure 4b. shows a line scan comparison between the topography, KPFM, Raman, and s-SNOM measurements. All the line scans were performed in proximity of the yellow dashed lines shown in Figures 2, 3b, and 4a.. The x-scales were translated to be approximately the same, based on topographic and optical imaging. It is important to mention that scans were not performed at the exact same location, so each scan has slightly different topography as shown in the figure. Starting with the KPFM, a slight potential difference of about 0.2 V is evident between the substrate and the drift layer. While the topography seems to be flat, the scan roughly shows a constant potential difference of 0.8V for the approximate length of the drift layer (8-micron) and then drops back to the 0.6V potential difference when entering the substrate region, where the n++ layer is barely observable. Moving on to the Raman scan done at the 720–760 cm$^{-1}$ spectral window, there is a clear drop in the normalized amplitude when crossing the drift layer to the substrate, and the signal changes over about a 2μm region upon entering the substrate. There is a relatively small change in the 760–950 cm$^{-1}$ spectral range, when moving from drift layer to the substrate. This shows the limitations of Raman mapping – it is unable to resolve stark changes in the carrier density well, which can be attributed to a combination of depth and spatial resolution. Finally, the s-SNOM results (at 115 cm$^{-1}$, 288 cm$^{-1}$, and 712 cm$^{-1}$) show variations in correlation with their s-SNOM mapping (Figure 4.) when moving from the drift layer to the substrate. A sharp decrease at the approximate interface of the substrate and the drift layer (at the proximity of -10μm position) is evident in all three scans. The scans also validate the approximate length of the epitaxial layer. As expected from mapping, the 288 cm$^{-1}$ scan show a gradient change in amplitude when moving away from the n++ interface in oppose to the sharp change in amplitude in 115 and 712cm$^{-1}$ scans. It is worth emphasizing that the manual polishing process inherently prevents the attainment of a perfectly flat surface which is why each scan (taken at different places) has a slightly different topography. Also, note that both Raman and s-SNOM line scan amplitude are normalized to 1. All in all, the comparisons show a higher overall sensitivity of the s-SNOM among all the other methods, showing the ability to resolve smaller features with clearer signals between different spatial regions.

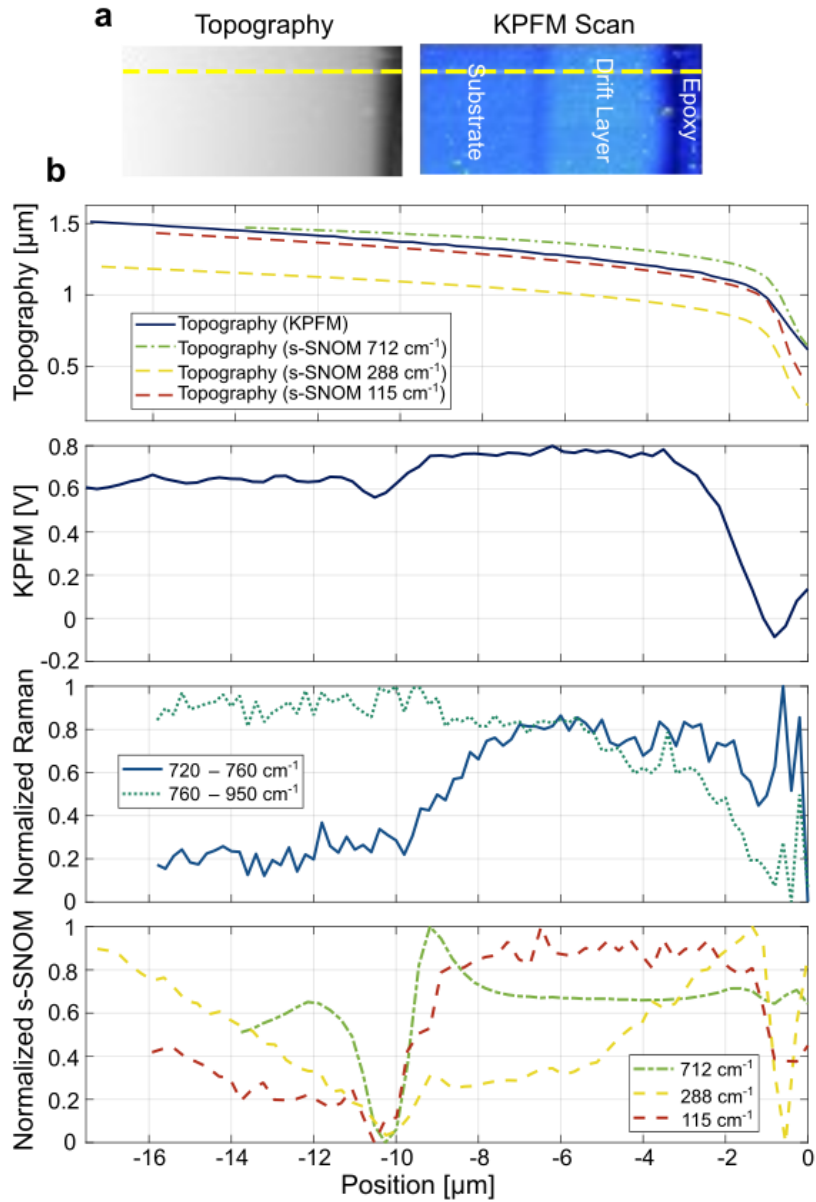

Figure 4: Side by side AFM topography and KPFM map of the sample, along with a line scan comparison of all methods mentioned so far. a) KPFM map of the GaN sample alongside the corresponding AFM scan covering a 20μmX10μm area. Note that the yellow dashed like represents the approximate position of the line scans shown in the bottom comparison. b) Stacked line scan comparison of information obtained through AFM topography, KPFM, Raman and s-SNOM.

## 4. Conclusion:

In this work, we have demonstrated that s-SNOM is an effective method for probing the internal structure of a GaN PIN diode. Our results show that combined THz and MIR imaging is particularly effective for understanding the presence of different material properties in devices at the nanoscale, due to the separate material signatures in both ranges. Specifically, THz allows us to examine only free carrier effects, whilst the MIR allows us to examine both free carrier and lattice effects. Analysis of the s-SNOM amplitude data revealed strong frequency-dependent contrast of different device layers with different carrier concentration at chosen wavenumbers that correlated with the simple point dipole model. Scans taken at 115cm$^{-1}$, 695cm$^{-1}$, and 712cm$^{-1}$ exhibited strong contrast suitable for identifying device structure as well as defects in the substrates, with extremely high spatial resolution not possible with other optical techniques. To validate our measurements, we conducted micro-Raman mapping and KPFM measurements. Raman mapping exhibited distinct spatial contrast between the substrate and the drift layers but also missed the small features such as the heavily dopped interface, nonuniformities across the substrate, and had broad peaks with high uncertainty for the carrier density. KPFM was also able to reveal a variation in surface potential difference between layers with different dopant concentration but cannot detect changes in the lattice. Overall, our results suggest broad bandwidth, multi-frequency s-SNOM imaging is uniquely capable of determining both carrier density and lattice variations within the GaN device. This study acts as a transferrable framework to other polar devices and demonstrates s-SNOM as a practical tool to examine variations in devices. Such information is invaluable in optimizing the performance of materials, as well as understanding failure methods.

**Acknowledgements**

Parts of this research were carried out at the ELBE Center for High-Power Radiation Sources at the Helmholtz- Zentrum Dresden-Rossendorf e.V., a member of the Helmholtz Association. under proposals 24103381-ST and 25103701-ST.

**Funding**

TGF and HZ acknowledge support from ONR Grant No. N00014-23-1-2616, "Ultrafast, nano-optic and temperature-dependent infrared (IR) probes for wide bandgap semiconductor characterization." J.D.C. and A. S. acknowledges the support of the Office of Naval Research under Grant #N00014-23-1-2676. K.D.G acknowledges funding from the Army Research Office under grant no. W911NF-21-1-0119. F.K., R.B., M.O. J.W., L.M.E., and S.C.K. acknowledge the financial support by the Bundesministerium für Bildung und Forschung (BMBF, Federal Ministry of Education and Research, Germany, Project Grant 05K19ODB, and 05K22ODA) and by the Deutsche Forschungsgemeinschaft (DFG, German Research Foundation) through the project CRC1415 as well as under Germany's Excellence Strategy through Würzburg-Dresden Cluster of Excellence on Complexity and Topology in Quantum Matter—ct.qmat (EXC 2147, project-id 390858490).

## References:

[1]     H. Fu, K. Fu, S. Chowdhury, T. Palacios, Y. Zhao, *IEEE Trans Electron Devices* **2021**, *68*.

[2]     H. Fu, K. Fu, S. Chowdhury, T. Palacios, Y. Zhao, *IEEE Trans Electron Devices* **2021**, *68*.

[3]     R. Sun, J. Lai, W. Chen, B. Zhang, *IEEE Access* **2020**, *8*.

[4]     A. Lidow, J. Strydom, M. de Rooij, D. Reusch, *GaN Transistors for Efficient Power Conversion: Second Edition*, Vol. 9781118844762, **2014**.

[5]     A. Udabe, I. Baraia-Etxaburu, D. G. Diez, *Gallium Nitride Power Devices: A State of the Art Review*, Vol. 11, **2023**.

[6]     S. Musumeci, V. Barba, *Energies (Basel)* **2023**, *16*.

[7]     B. J. Baliga, *Gallium nitride devices for power electronic applications*, Vol. 28, **2013**.

[8]     S. N. Mohammad, A. A. Salvador, H. Morkoç, *Proceedings of the IEEE* **1995**, *83*.

[9]     X. Ding, Y. Zhou, J. Cheng, *CES Transactions on Electrical Machines and Systems* **2019**, *3*.

[10]    M. N. Yoder, In *Proceedings of the IEEE Cornell Conference on Advanced Concepts in High Speed Semiconductor Devices and Circuits*, **1997**.

[11]    R. F. Davis, Z. Sitar, B. E. Williams, H. S. Kong, H. J. Kim, J. W. Palmour, J. A. Edmond, J. Ryu, J. T. Glass, C. H. Carter, *Materials Science and Engineering B* **1988**, *1*.

[12]    H. Shibata, Y. Waseda, H. Ohta, K. Kiyomi, K. Shimoyama, K. Fujito, H. Nagaoka, Y. Kagamitani, R. Simura, T. Fukuda, In *Materials Transactions*, **2007**.

[13]    F. Medjdoub, K. Iniewski, *Gallium nitride (GaN): Physics, devices, and technology*, **2017**.

[14]    J. C. Gallagher, M. A. Ebrish, M. A. Porter, A. G. Jacobs, B. P. Gunning, R. J. Kaplar, K. D. Hobart, T. J. Anderson, *Sci Rep* **2022**, *12*.

[15]    P. Boguslawski, E. L. Briggs, J. Bernholc, *Phys Rev B* **1995**, *51*.

[16]    L. Liu, J. H. Edgar, *Substrates for gallium nitride epitaxy*, Vol. 37, **2002**.

[17]    G. Meneghesso, M. Meneghini, A. Tazzoli, N. Ronchi, A. Stocco, A. Chini, E. Zanoni, *Int J Microw Wirel Technol* **2010**, *2*.

[18]    M. A. Berkson, E. A. Pogue, M. E. Bartlett, S. A. Shuler, M. M. Kesavan, T. J. Montalbano, A. L. Bennett-Jackson, J. B. Abraham, I. Z. Martins, T. Terlier, J. C. Gagnon, *Cryst Growth Des* **2024**, *24*, 3149.

[19]    N. Stoddard, S. Pimputkar, *Progress in Ammonothermal Crystal Growth of Gallium Nitride from 2017–2023: Process, Defects and Devices*, Vol. 13, **2023**.

[20]    T. Sochacki, L. Kirste, K. Sierakowski, A. Jaroszyńska, R. Jakieła, M. Fijałkowski, K. Grabiańska, M. Zając, J. S. Koziorowska, A. Lachowski, M. Turek, P. Straňák, K. Sumida, M. Boćkowski, *Appl Surf Sci* **2025**, *699*.

[21]    R. Chierchia, T. Böttcher, H. Heinke, S. Einfeldt, S. Figge, D. Hommel, *J Appl Phys* **2003**, *93*.

[22]    P. Li, Y. Liu, X. Meng, *Revista Mexicana de Fisica* **2016**, *62*.

[23]    P. Senthil Kumar, K. Grace Pavithra, M. Naushad, In *Nanomaterials for Solar Cell Applications*, **2019**.

[24]    K. Gruszka, M. Nabiałek, K. Błoch, S. Walters, *International Journal of Materials Research* **2015**, *106*.

[25]    A. Sakai, H. Sunakawa, A. Usui, *Appl Phys Lett* **1998**, *73*.

[26]    F. A. Ponce, D. Cherns, W. T. Young, J. W. Steeds, *Appl Phys Lett* **1996**, *69*.

[27]    S. Christiansen, M. Albrecht, H. P. Strunk, *Philosophical Magazine A: Physics of Condensed Matter, Structure, Defects and Mechanical Properties* **1998**, *77*.

[28]    Y. Yao, K. Sato, Y. Sugawara, Y. Ishikawa, *J Alloys Compd* **2022**, *902*.

[29]     H. A. A. Abdul Amir, M. A. Fakhri, A. Abdulkhaleq Alwahib, In *Materials Today: Proceedings*, **2021**.

[30]     B. Hauer, C. E. Marvinney, M. Lewin, N. A. Mahadik, J. K. Hite, N. Bassim, A. J. Giles, R. E. Stahlbush, J. D. Caldwell, T. Taubner, *Adv Funct Mater* **2020**, *30*.

[31]     T. Nakamura, N. Ishida, K. Sagisaka, Y. Koide, *AIP Adv* **2020**, *10*.

[32]     F. Huth, A. Govyadinov, S. Amarie, W. Nuansing, F. Keilmann, R. Hillenbrand, *Nano Lett* **2012**, *12*.

[33]     A. J. Giles, S. Dai, O. J. Glembocki, A. V. Kretinin, Z. Sun, C. T. Ellis, J. G. Tischler, T. Taniguchi, K. Watanabe, M. M. Fogler, K. S. Novoselov, D. N. Basov, J. D. Caldwell, *Nano Lett* **2016**, *16*.

[34]     A. Fali, S. T. White, T. G. Folland, M. He, N. A. Aghamiri, S. Liu, J. H. Edgar, J. D. Caldwell, R. F. Haglund, Y. Abate, *Nano Lett* **2019**, *19*.

[35]     W. Melitz, J. Shen, A. C. Kummel, S. Lee, *Kelvin probe force microscopy and its application*, Vol. 66, **2011**.

[36]     A. Kaschner, A. Hoffmann, C. Thomsen, F. Bertram, T. Riemann, J. Christen, K. Hiramatsu, H. Sone, N. Sawaki, *Appl Phys Lett* **2000**, *76*.

[37]     M. S. Mohajerani, S. Khachadorian, T. Schimpke, C. Nenstiel, J. Hartmann, J. Ledig, A. Avramescu, M. Strassburg, A. Hoffmann, A. Waag, *Appl Phys Lett* **2016**, *108*.

[38]     T. Nakamura, N. Ishida, K. Sagisaka, Y. Koide, *AIP Adv* **2020**, *10*.

[39]     V. V. Strelchuk, A. S. Nikolenko, P. M. Lytvyn, A. S. Romanyuk, Y. I. Mazur, M. E. Ware, E. A. Decuir, G. J. Salamo, A. E. Belyaev, *Physica Status Solidi (C) Current Topics in Solid State Physics* **2014**, *11*.

[40]     N. Wieser, M. Klose, R. Dassow, F. Scholz, J. Off, *J Cryst Growth* **1998**, *189–190*.

[41]     P. Perlin, J. Camassel, W. Knap, T. Taliercio, J. C. Chervin, T. Suski, I. Grzegory, S. Porowski, *Appl Phys Lett* **1995**, *67*.

[42]     F. Kuschewski, H. G. Von Ribbeck, J. Döring, S. Winnerl, L. M. Eng, S. C. Kehr, *Appl Phys Lett* **2016**, *108*.

[43]     A. Kasic, M. Schubert, S. Einfeldt, D. Hommel, T. E. Tiwald, *Phys Rev B Condens Matter Mater Phys* **2000**, *62*.

[44]     T. Taubner, F. Keilmann, R. Hillenbrand, *Opt Express* **2005**, *13*.

[45]     T. Taubner, D. Korobkin, Y. Urzhumov, G. Shvets, R. Hillenbrand, *Science (1979)* **2006**, *313*.

[46]     X. Chen, D. Hu, R. Mescall, G. You, D. N. Basov, Q. Dai, M. Liu, *Modern Scattering-Type Scanning Near-Field Optical Microscopy for Advanced Material Research*, Vol. 31, **2019**.

[47]     T. V. A. G. de Oliveira, T. Nörenberg, G. Álvarez-Pérez, L. Wehmeier, J. Taboada-Gutiérrez, M. Obst, F. Hempel, E. J. H. Lee, J. M. Klopf, I. Errea, A. Y. Nikitin, S. C. Kehr, P. Alonso-González, L. M. Eng, *Advanced Materials* **2021**, *33*.

[48]     T. Nörenberg, G. Álvarez-Pérez, M. Obst, L. Wehmeier, F. Hempel, J. M. Klopf, A. Y. Nikitin, S. C. Kehr, L. M. Eng, P. Alonso-González, T. V. A. G. De Oliveira, *ACS Nano* **2022**, *16*.

[49]     L. Novotny, *Chapter 5 The history of near-field optics*, Vol. 50, **2007**.

[50]     M. He, T. G. Folland, J. Duan, P. Alonso-González, S. De Liberato, A. Paarmann, J. D. Caldwell, *Anisotropy and Modal Hybridization in Infrared Nanophotonics Using Low-Symmetry Materials*, Vol. 9, **2022**.

[51]     L. V. Brown, M. Davanco, Z. Sun, A. Kretinin, Y. Chen, J. R. Matson, I. Vurgaftman, N. Sharac, A. J. Giles, M. M. Fogler, T. Taniguchi, K. Watanabe, K. S. Novoselov, S. A. Maier, A. Centrone, J. D. Caldwell, *Nano Lett* **2018**, *18*.

[52]     T. G. Folland, L. Nordin, D. Wasserman, J. D. Caldwell, *J Appl Phys* **2019**, *125*.

[53]     A. Centrone, *Infrared Imaging and Spectroscopy beyond the Diffraction Limit∗*, Vol. 8, **2015**.

[54]     B. Knoll, F. Keilmann, *Opt Commun* **2000**, *182*.

[55]     R. Hillenbrand, B. Knoll, F. Keilmann, In *Journal of Microscopy*, **2001**.

[56]     L. Wehmeier, D. Lang, Y. Liu, X. Zhang, S. Winnerl, L. M. Eng, S. C. Kehr, *Phys Rev B* **2019**, *100*.

[57]     M. Lucidi, D. E. Tranca, L. Nichele, D. Ünay, G. A. Stanciu, P. Visca, A. M. Holban, R. Hristu, G. Cincotti, S. G. Stanciu, *Gigascience* **2020**, *9*.

[58]     A. M. Gigler, A. J. Huber, M. Bauer, A. Ziegler, R. Hillenbrand, R. W. Stark, *Opt Express* **2009**, *17*.

[59]     W. Mao, H. Wang, P. Shi, C. Yang, Y. Zhang, X. Zheng, C. Wang, J. Zhang, Y. Hao, *Semicond Sci Technol* **2018**, *33*.

[60]     H. Zandipour, M. Madani, *Journal of Semiconductors* **2020**, *41*.

[61]     T. Zhu, R. A. Oliver, *Unintentional doping in GaN*, Vol. 14, **2012**.

[62]     H. Zandipour, M. Mohammad R, *Secured floating gate transistor and method for securing floating gate transistors*, **2021**.

[63]     N. A. Aghamiri, F. Huth, A. J. Huber, A. Fali, R. Hillenbrand, Y. Abate, *Opt Express* **2019**, *27*.

[64]     P. Pandey, T. M. Nelson, W. M. Collings, M. R. Hontz, D. G. Georgiev, A. D. Koehler, T. J. Anderson, J. C. Gallagher, G. M. Foster, A. Jacobs, M. A. Ebrish, B. P. Gunning, R. J. Kaplar, K. D. Hobart, R. Khanna, *IEEE Trans Electron Devices* **2022**, *69*, 5096.

[65]     T. Vincent, X. Liu, D. Johnson, L. Mester, N. Huang, O. Kazakova, R. Hillenbrand, J. L. Boland, **2024**.

[66]     M. Helm, S. Winnerl, A. Pashkin, J. M. Klopf, J. C. Deinert, S. Kovalev, P. Evtushenko, U. Lehnert, R. Xiang, A. Arnold, A. Wagner, S. M. Schmidt, U. Schramm, T. Cowan, P. Michel, *Eur Phys J Plus* **2023**, *138*.

[67]     F. H. Feres, R. A. Mayer, L. Wehmeier, F. C. B. Maia, E. R. Viana, A. Malachias, H. A. Bechtel, J. M. Klopf, L. M. Eng, S. C. Kehr, J. C. González, R. O. Freitas, I. D. Barcelos, *Nat Commun* **2021**, *12*.

[68]     V. Janonis, A. Cernescu, P. Prystawko, R. Januškevičius, S. Indrišiūnas, I. Kašalynas, *Materials* **2025**, *18*.

[69]     J. D. Caldwell, O. J. Glembocki, Y. Francescato, N. Sharac, V. Giannini, F. J. Bezares, J. P. Long, J. C. Owrutsky, I. Vurgaftman, J. G. Tischler, V. D. Wheeler, N. D. Bassim, L. M. Shirey, R. Kasica, S. A. Maier, *Nano Lett* **2013**, *13*.

[70]     S. G. Criswell, N. A. Mahadik, J. C. Gallagher, J. Barnett, L. Kim, M. Ghorbani, B. Kamaliya, N. D. Bassim, T. Taubner, J. D. Caldwell, *Nano Lett* **2024**, *24*.

[71]     J. D. Caldwell, L. Lindsay, V. Giannini, I. Vurgaftman, T. L. Reinecke, S. A. Maier, O. J. Glembocki, *Low-loss, infrared and terahertz nanophotonics using surface phonon polaritons*, Vol. 4, **2015**.

[72]     Y. Zeng, J. Ning, J. Zhang, Y. Jia, C. Yan, B. Wang, D. Wang, *Applied Sciences (Switzerland)* **2020**, *10*.

[73]     H. Harima, *Properties of GaN and related compounds studied by means of Raman scattering*, Vol. 14, **2002**.

[74]     J. C. Burton, L. Sun, M. Pophristic, S. J. Lukacs, F. H. Long, Z. C. Feng, I. T. Ferguson, *J Appl Phys* **1998**, *84*, 6268.

[75]     N. Armakavicius, S. Knight, P. Kühne, V. Stanishev, D. Q. Tran, S. Richter, A. Papamichail, M. Stokey, P. Sorensen, U. Kilic, M. Schubert, P. P. Paskov, V. Darakchieva, *APL Mater* **2024**, *12*.